# Plasma Upflows Induced by Magnetic Reconnection Above an Eruptive Flux Rope

Deborah Baker[1] · Teodora Mihailescu[1] · Pascal Démoulin[2] · Lucie M. Green[1] ·
Lidia van Driel-Gesztelyi[1,2,3] · Gherardo Valori[1,4] · David H. Brooks[5] ·
David M. Long[1] · Miho Janvier[6]

**Abstract**
One of the major discoveries of *Hinode's Extreme-ultraviolet Imaging Spectrometer* (EIS) is the presence of upflows at the edges of active regions. As active regions are magnetically connected to the large-scale field of the corona, these upflows are a likely contributor to the global mass cycle in the corona. Here we examine the driving mechanism(s) of the very strong upflows with velocities in excess of 70 km s$^{-1}$, known as blue-wing asymmetries, observed during the eruption of a flux rope in AR 10977 (eruptive flare SOL2007-12-07T04:50). We use *Hinode*/EIS spectroscopic observations combined with magnetic-field modeling to investigate the possible link between the magnetic topology of the active region and the strong upflows. A Potential Field Source Surface (PFSS) extrapolation of the large-scale field shows a quadrupolar configuration with a separator lying above the flux rope. Field lines formed by induced reconnection along the separator before and during the flux-rope eruption are spatially linked to the strongest blue-wing asymmetries in the upflow regions. The flows are driven by the pressure gradient created when the dense and hot arcade loops of the active region reconnect with the extended and tenuous loops overlying it. In view of the fact that separator reconnection is a specific form of the more general quasi-separatrix (QSL) reconnection, we conclude that the mechanism driving the strongest upflows is, in fact, the same as the one driving the persistent upflows of $\approx 10-20$ km s$^{-1}$ observed in all active regions.

✉ D. Baker
deborah.baker@ucl.ac.uk

[1] Mullard Space Science Laboratory, University College London, Holmbury St. Mary, Dorking, Surrey, RH5 6NT, UK

[2] LESIA, Observatoire de Paris, Université PSL, CNRS, Sorbonne Université, Univ. Paris Diderot, Sorbonne Paris Cité, 5 place Jules Janssen, 92195 Meudon, France

[3] Konkoly Observatory, Research Centre for Astronomy and Earth Sciences, Konkoly Thege út 15-17., 1121, Budapest, Hungary

[4] Max Planck Institute for Solar System Research, Justus-von-Liebig-Weg 3, 37077 Göttingen, Germany

[5] College of Science, George Mason University, 4400 University Drive, Fairfax, VA 22030, USA

[6] Université Paris-Saclay, CNRS, Institut d'Astrophysique Spatiale, Orsay, France

## 1. Introduction

Spectroscopic observations dating back to *Skylab* in the 1970s have revealed the presence of blue-shifted plasma upflows in the quiet Sun, coronal holes, and active regions. When the plasma lies on open magnetic field then upflows become outflows into the solar wind, however upflows can also be confined along closed loops. This is an important distinction in that upflows include outflows but not all upflows become outflows into the heliosphere. Plasma upflows are characteristic of many active regions, occurring towards the boundary where they impact (and are potentially formed by) the surrounding environment on both small and large scales. They appear to play some role in almost all aspects of mass and energy supply to the active-region corona and potentially the solar wind. They may be a signature of mass circulation (Marsch et al., 2008; McIntosh et al., 2012), connection to distant regions (Boutry et al., 2012), or evidence of outflow into the slow wind (Brooks, Ugarte-Urra, and Warren, 2015). Details of the upflow generation are still debated, in particular whether they are driven from low in the atmosphere, or form as a response to magnetic topological reorganization higher up (Polito et al., 2020), but closed/open field changes at the active-region periphery are always involved, which invokes some form of magnetic reconnection. There is evidence that such processes are occurring in the upflows from radio noise storms detected on the ground (Del Zanna et al., 2011; Mandrini et al., 2015), close to the Sun from *Parker Solar Probe* (Harra et al., 2021), and from studies of the sources of solar energetic particles (Brooks and Yardley, 2021). It is therefore important to understand the formation mechanisms of the upflows.

### 1.1. General Characteristics of Active Region Upflows

Active-region upflows are distinct from those plasma flows originating from coronal holes and the quiet Sun, e.g. intermittent jets, spicules, and surges. Recent spectroscopic studies based on observations from *Hinode's EUV Imaging Spectrometer* (EIS: Culhane et al., 2007) have provided an overview of their general characteristics. These large-scale and steady bulk plasma flows are localized at the following and leading sides of active regions with each upflow area associated with a magnetic monopolar field, which can be a sunspot or multiple flux tubes of the same polarity (more typical for the following polarity). Upflows are observed to occur in pairs, whether as a single pair in isolated bipolar active regions, or as multiple pairs in more complex configurations (Baker et al., 2017). The magnitude of the velocities observed in strong coronal emission lines, such as Fe XII 195.12 Å and Fe XIII 202.04 Å, is in the range of [5, 50] km s$^{-1}$ when fitted with single Gaussian functions (e.g. Harra et al., 2008; Doschek et al., 2008; Del Zanna, 2008; Baker et al., 2009; Tian et al., 2021). In a few active regions, the blue wing is much more extended than the red wing, so that significant blue-wing asymmetries of more than 100 km s$^{-1}$ have been observed (e.g. Hara et al., 2008; De Pontieu et al., 2009; Bryans, Young, and Doschek, 2010; Peter, 2010; Tian, McIntosh, and De Pontieu, 2011; Tian et al., 2011; Doschek, 2012; Brooks and Warren, 2012; Tian et al., 2021, and the references therein). The possible driving mechanisms of these high-speed upflows are not well understood.

Upflows first appear during the emergence phase (Harra et al., 2012), and they persist for the time that active regions are observed on disk (Démoulin et al., 2013; Baker et al., 2017; Tian et al., 2021) and for multiple solar rotations (Zangrilli and Poletto, 2016; Harra et al., 2017). When continually tracked during disk transit, there is a clear evolution of the upflows that is related to the solar rotation progressively changing the viewpoint of the flows. Such an evolution indicates the presence of a strong and highly collimated component to the upflows that peaks when the collimated component is parallel to the line of sight. Démoulin et al. (2013) modeled this behavior in order to separate rotation-related apparent evolution from intrinsic evolution due to flux emergence or coronal activity such as flaring and coronal mass ejections (CMEs). From the stationary-flow model, the 3D structure of the large-scale upflows can be deduced. The flows are thin, fan-like structures that are tilted away from the active-region center in agreement with the configuration determined from magnetic-field extrapolations (Démoulin et al., 2013; Baker et al., 2017). Démoulin et al. (2013) concluded that the same flows are observed along magnetic-field lines in spectral lines of different temperatures, but due to the thermal stratification of the solar atmosphere, the spatial locations and extents of the flows are different.

### 1.2. Possible Driving Mechanisms

Possible driving mechanisms of active-region upflows must account for their main characteristics observed in all active regions, the most challenging of which are their longevity, spatial stability, and occurrence in pairs. First proposed by Baker et al. (2009), magnetic reconnection along quasi-separatrix layers (QSLs) is one driving mechanism that can provide a framework for these observed global properties of active-region upflows (Baker et al., 2009; van Driel-Gesztelyi et al., 2012; Démoulin et al., 2013; Baker et al., 2017). QSLs are thin 3D volumes where magnetic fields display strong gradients in magnetic connectivity and they are preferential sites for current sheet formation and magnetic reconnection (Démoulin et al., 1996; Démoulin, 2006). QSLs are present in active regions where internally connecting loops and externally connected/open loops are rooted, forming strong connectivity gradients on either side of active regions. The low-corona footprint of such QSLs are located over each peripheral magnetic polarity of any active region. These are the precise locations of upflows naturally occuring in pairs over leading and following polarities. Furthermore, QSLs are defined by the global properties of the large-scale photospheric magnetic-flux distribution that evolves slowly, as compared to typical coronal time scales, thereby resulting in locations of sustained/successive reconnection capable of driving upflows for days and weeks. In the reconnected loops, upflows are physically driven along magnetic field by the pressure gradient created when the short loops at the periphery of an active region's core reconnect with the long externally connected loops of lower density rooted in the same polarity (Baker et al., 2009; Bradshaw, Aulanier, and Del Zanna, 2011; Del Zanna et al., 2011; van Driel-Gesztelyi et al., 2012; Démoulin et al., 2013; Mandrini et al., 2015; Baker et al., 2017).

QSL reconnection has been invoked to be at the origin of coronal upflows, however Polito et al. (2020) have recently shown that weakly blue-shifted plasma is observed in the lower solar atmosphere in the same spatial locations as the coronal upflows discussed so far. In their study, small-scale upflows in the chromospheric C II and Mg II k2 lines and the transition region Si IV line were detected by the *Interface Region Imaging Spectrometer* (IRIS: De Pontieu et al., 2014). QSL reconnection is a plausible driving mechanism for lower-temperature upflows (e.g. Démoulin et al., 2013) although this has not been demonstrated to date. For the transition-region and coronal lines observed with EIS, detailed analysis complemented by magnetic modeling (Baker et al., 2009; Démoulin et al., 2013) has demonstrated that the active-region upflows in the hotter lines, e.g. Fe XII and Fe XV, and the downflows in the cooler Si VII line have the same structure, are roughly co-spatial, and originate at QSLs. It is not clear, however, and will require further analysis of the flows in chromospheric IRIS lines, whether or not the weak upflows in Polito et al. (2020) are similarly structured by the magnetic topology and indeed linked to active-region upflows.

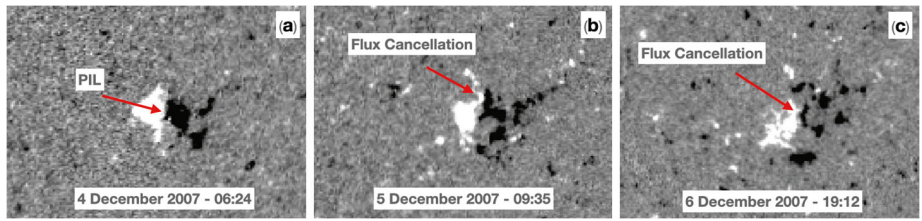

**Figure 1** SOHO/MDI line of sight magnetograms. Locations of flux cancelation along the main polarity inversion line (PIL) are indicated by red arrows. Spatial scale for each image is $X = 350''$, $Y = 250''$.

### 1.3. Article Road Map

Here we present *Hinode*/EIS observations of AR 10977 that capture the evolution of blueshifted upflows prior to, during, and after a failed eruption that is then followed by a successful eruption with a CME. All velocity parameters, including blue-wing asymmetries, increase before and during the eruptive events then decrease throughout the time of their decay phases. Magnetic field modeling is employed to interpret the evolution of the upflows in the context of the 3D magnetic topology and reconnection. The article is structured as follows: Section 2 summarizes the key features of the development and evolution of the sigmoid/flux rope from Green, Kliem, and Wallace (2011). This is followed by a full account of the *Hinode*/EIS observations and methods in Section 3. The upflow evolution is described and quantified in Section 4. In Section 5 we proffer our interpretation of these events before concluding in Section 6.

## 2. Sigmoidal Active Region – AR 10977

### 2.1. Evolution of Photospheric Magnetic Field

For a full account of the photospheric and coronal evolution of the active region from emergence to decay, we refer the reader to Green, Kliem, and Wallace (2011). Here, we provide a summary of the key elements relevant to the formation of a sigmoid that erupts as a flux rope/CME on 7 December 2007. AR 10977 appeared at the east limb on 1 December 2007. For the next three days, the photospheric magnetic-field evolution was dominated by flux emergence until $\approx$ 06:00 UT on 4 December when a peak flux of $\approx 2 \times 10^{21}$ Mx was reached. Flux cancelation prevailed during the active region's decay phase. Figure 1 shows a series of magnetograms from the *Michelson Doppler Imager* (MDI) onboard the *Solar and Heliospheric Observatory* (SOHO) beginning around the time of the peak flux and continuing for the early stages of the decay phase. In panel a, the red arrow identifies the main or internal polarity inversion line (PIL) in between the still coherent positive and negative polarities. Later on, the primary site of flux cancelation is in the northern section of the PIL as indicated by the red arrows in panels b and c. Approximately one-third of the peak flux was canceled at this location over 2.5 days prior to a CME early on 7 December (Green, Kliem, and Wallace, 2011).

**Table 1** EIS study details for AR 10977. $X$- and $Y$-locations are adjusted for alignment with SOHO/ EIT and MDI. All dates are December 2007 and times are UT.

| Date | Start time | $X$ [″] | $Y$ [″] | Study No. | Exposure duration [s] | FoV [″] | Raster duration [min] |
|---|---|---|---|---|---|---|---|
| 6 Dec | 12:03 | −82 | −152 | 180 | 50 | 180×512 | 50 |
| 7 Dec | 00:18 | 34 | −150 | 180 | 50 | 180×512 | 50 |
| 7 Dec | 01:15 | 40 | −150 | 45 | 60 | 128×512 | 128 |
| 7 Dec | 03:27 | 63 | −148 | 180 | 50 | 180×512 | 50 |
| 7 Dec | 06:37 | 93 | −148 | 180 | 50 | 180×512 | 50 |
| 7 Dec | 11:26 | 141 | −147 | 180 | 50 | 180×512 | 50 |

### 2.2. Sigmoid Formation and Eruption

Figure 2 contains a series of *Solar Terrestrial Relations Observatory–A* (STEREO–A) *Extreme Ultraviolet Imager* (EUVI) 284 Å images processed using the Multi-Scale Gaussian Normalization (MGN) technique of Morgan and Druckmüller (2014) from the animation included in the Supplementary Information showing the evolution of the coronal field from 23:46 UT on 5 December to 05:46 UT on 7 December. Concurrent with the continuous flux cancelation and dispersion of the negative photospheric magnetic field, the coronal arcade field became more and more sheared in the northern region while remaining near potential in the southern part of the active region. By 16:00 UT on 6 December, a continuous forward *S*-sigmoid is evident in the EUV images (see panel b). The sigmoid center is located above the primary site of flux cancelation in the SOHO/MDI magnetograms of Figure 1. Green, Kliem, and Wallace (2011) reported that the forward *S*-structure was observed in soft X-ray images over the same time period. The combination of significant flux cancelation within a sheared arcade followed by the appearance of a sigmoid suggests that a flux-rope structure had formed in the active region prior to the eruptions (van Ballegooijen and Martens, 1989; Green, Kliem, and Wallace, 2011).

The sigmoid/flux-rope structure continued to evolve before it finally erupted on 7 December (SOL2007-12-07T04:50). The eruption occurred in two stages, the approximate timings of which are indicated by the shaded regions overlaid on the *Hinode*/XRT soft X-ray (C poly) light curve in Figure 3. First, a failed eruption (blue region) began just after 00:00 UT. Figure 2 d shows the sigmoid during this stage at 02:06 UT as it is expanding/rising in the North. Second, the CME (yellow region) began just after 04:00 UT and was associated with a GOES Class B1.4 flare and a global wave (see Green, Kliem, and Wallace, 2011, and the references therein). STEREO–A observes the CME in panel e of Figure 2. Approximately one hour after the CME, the sigmoid structure was transformed to an arcade field. The post-eruption arcade loops are clearly visible in panel f.

### 3. *Hinode*/EIS Observations

#### 3.1. Data Reduction

*Hinode*/EIS observed AR 10977 on 6–7 December 2007 during the period in which the sigmoid/magnetic flux rope formed and erupted as a CME. The details of the EIS datasets are provided in Table 1. They were acquired using the 1″ slit in the normal scanning mode for study #45 and the sparse scanning mode for study #180 ($3''$ step size). The fields of view (FOV) were created by scanning $128''$–$180''$ in the solar-$X$ direction with each raster having a height of $512''$ in the solar-$Y$ direction. All EIS data were processed using the standard eis_prep routine available in the *Solar Soft* library (SSW: Freeland and Handy, 1998). This routine converts the measured CCD signal into calibrated intensity units of $\text{erg}\,\text{cm}^{-2}\,\text{s}^{-1}\,\text{sr}^{-1}\,\text{Å}^{-1}$. It accounts for the CCD dark current and pedestal, and treats dusty, hot, warm pixels, and cosmic rays. Orbital drift of the EIS spectra was removed using the neural network model of Kamio et al. (2010).

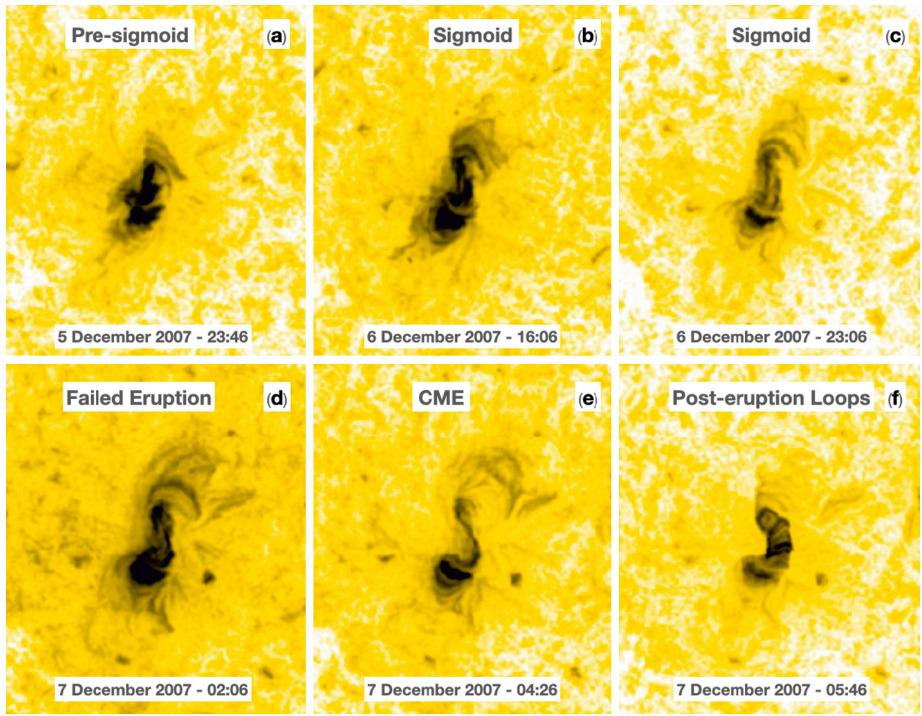

**Figure 2** STEREO–A 284 Å reverse color table images showing the coronal evolution of AR 10977 before, during, and after the failed and CME (=successful) eruptions. Images are processed using the MGN processing technique of Morgan and Druckmüller (2014). Spatial scale for each image is $X = 350''$, $Y = 400''$. (Included as animation called Fig2.mp4).

### 3.2. *Hinode*/EIS Doppler and Nonthermal Velocities

The Fe XII 195.12 Å and the Fe XIII 202.04 Å spectral lines were used in this study of upflows in AR 10977. A double-Gaussian function was fitted to the Fe XII 195.12 Å line to remove the self-blend at 195.18 Å. We used the method of Young et al. (2009) where the secondary component at 195.18 Å is forced to be 0.06 Å to the long-wavelength side of the main component at 195.12 Å and the width of the 195.18 Å line is set to be equal to that of the main component. The free parameters for the 195.12 Å line are the line peak, centroid, and width and for the 195.18 Å line only the line peak is free to vary. An example of the double-Gaussian fit from the upflow region at 11.26 UT on 7 December 2007 is shown in Figure 4. The effect of the blend is sensitive to density and becomes relatively strong for densities greater than $\log N_e = 10$ cm$^{-3}$ (Young, O'Dwyer, and Mason, 2012). We expect the effect to be minimal since the mean density within the upflow regions for all rasters used in this study is less than $\log N_e = 9$ cm$^{-3}$. For a detailed analysis of the Fe XII 195.12 line, see section 5.2 and Appendix A of Young, O'Dwyer, and Mason (2012). A single-Gaussian function was fitted to the unblended Fe XIII line.

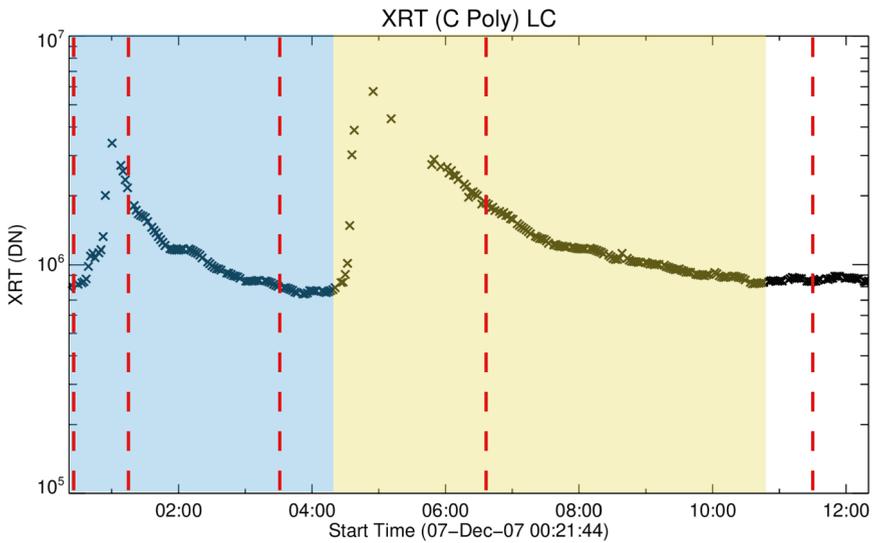

**Figure 3** *Hinode*/XRT soft X-ray (C poly) light curve of the entire active region covering the period from 00:21 UT to 12:15 UT on 7 December 2007. Blue/yellow shaded regions show the time range of the failed/CME (=successful) eruptions. *Hinode*/EIS raster times are indicated by the dashed red lines (00:18, 01:15, 03:27, 06:37, and 11:26 UT).

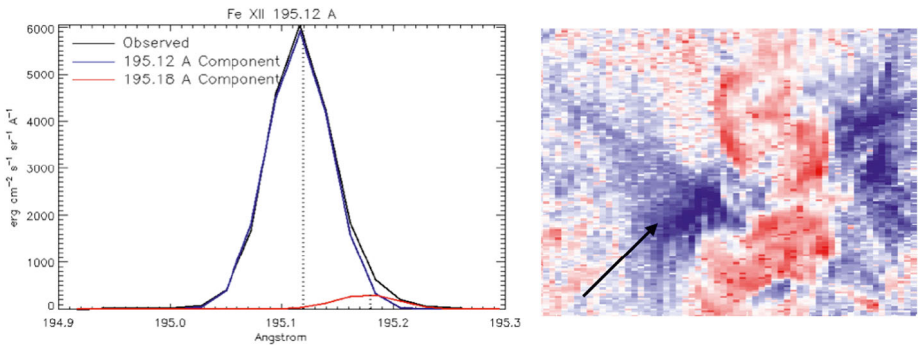

**Figure 4** Left: Example of a double-Gaussian fit to the Fe XII 195.12 Å emission line in the upflow region of the raster at 11:26 UT on 7 December 2007. The wavelength difference between the main and secondary components is fixed at 0.06 Å and the width of the 195.18 Å component is set to be equal to that of the main component (Young et al., 2009). Right: Arrow indicates location of the pixel of the profile (left) in the Fe XII Doppler velocity map at 11:26 UT. The image covers $180'' \times 120''$.

*Hinode*/EIS does not have an absolute wavelength calibration therefore relative Doppler velocities must be measured versus a reference wavelength. Since the FOV is dominated by quiet-Sun regions to the North ($Y = [0, 100]''$) and South ($Y = [-150, -390]''$) of the active region, we determined the reference wavelength for each emission line by averaging the centroid wavelengths of all pixels within the data array of each raster (i.e. for the full length in $Y$ of 512 pixels). Depending on the region and the raster, there is up to $\approx 3$ km s$^{-1}$ difference in the reference wavelength if a quiet-Sun patch is used instead of averaging over all centroid wavelengths in the FOV.

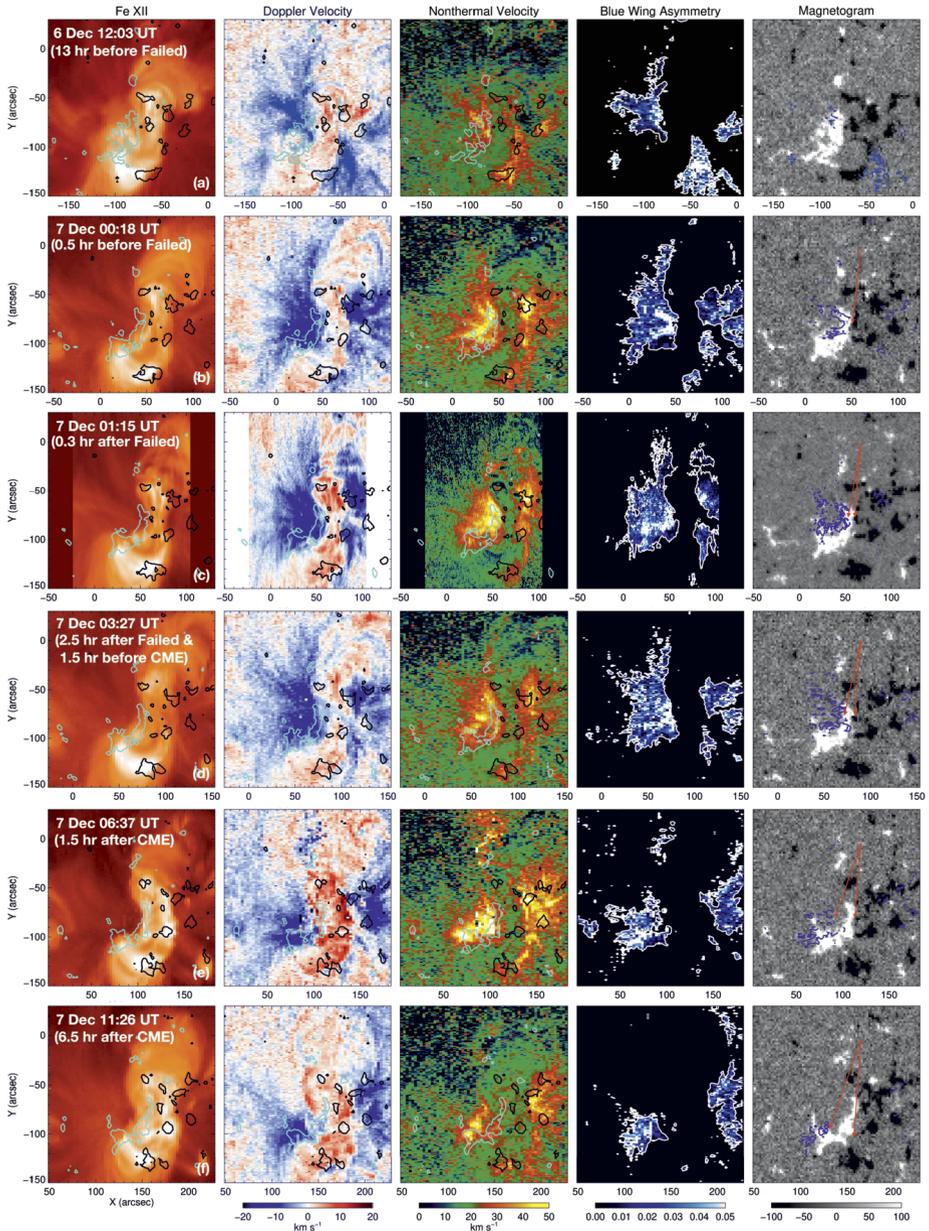

**Figure 5** Left to right: *Hinode*/EIS Fe XII 195.12 Å intensity, Doppler velocity, and nonthermal velocity maps overlaid with MDI contours of ±100 G (cyan/black), blue-wing asymmetry ($\alpha_{\mathrm{asym}}$) maps overplotted with Doppler velocity contour of –5 km s$^{-1}$ (white), and magnetogram overplotted with $\alpha_{\mathrm{asym}}$ contour of 0.05. Asymmetries maps are saturated at 0.05. All Fe XII 195.12 Å maps have had the self-blend at 195.18 Å removed. Notation in the first column corresponds to Figures 3 and 8. "Failed" refers to the failed eruption and CME to the CME event. Red arrows indicate the locations of the PIL and the western edge of the $\alpha_{\mathrm{asym}}$ contours as they separate at 00:18 UT – 06:37 UT.

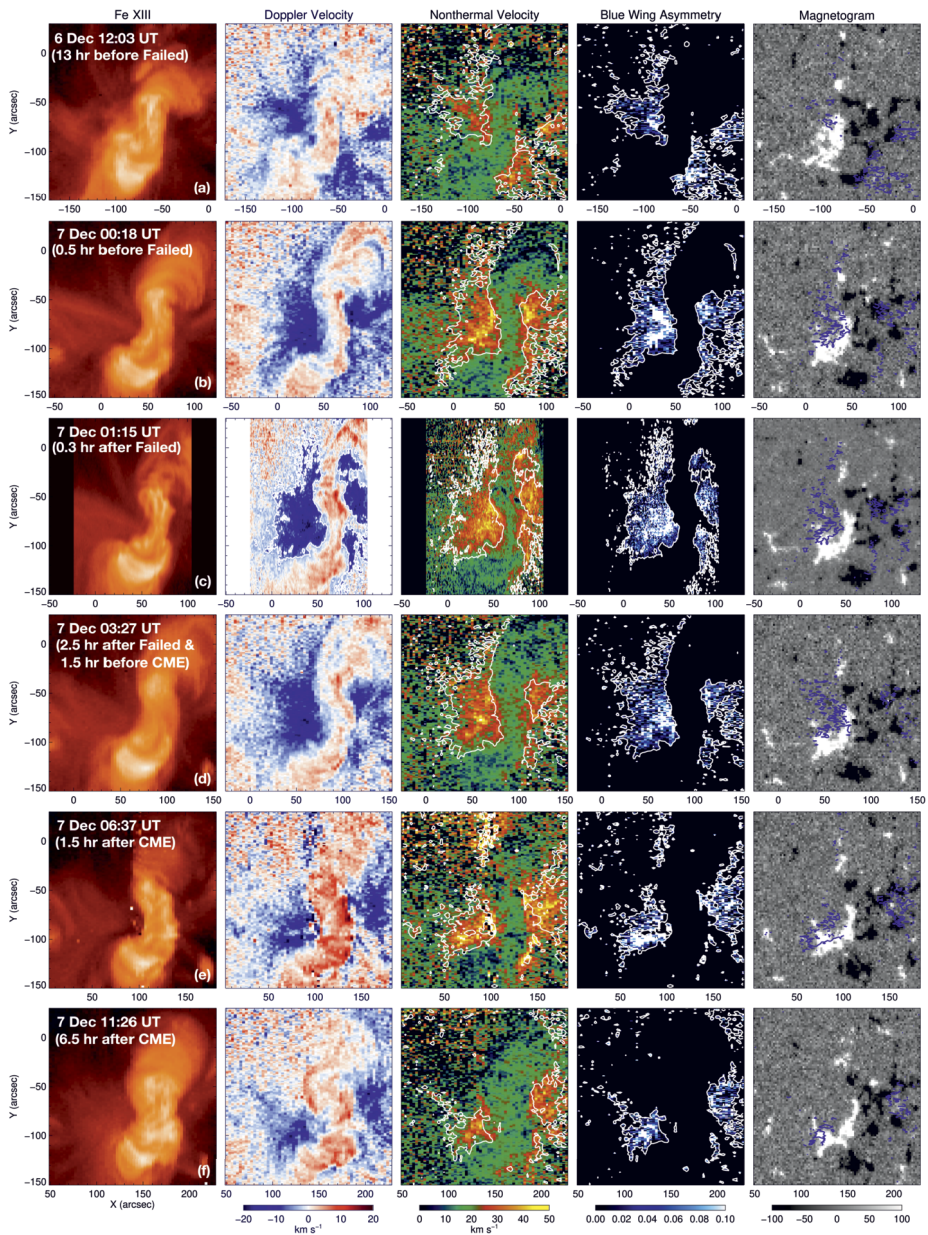

**Figure 6** Left to right: *Hinode*/EIS single-Gaussian Fe XIII 202.04 Å intensity, Doppler velocity, nonthermal velocity, blue-wing asymmetry ($\alpha_{\text{asym}}$) maps, and MDI magnetogram. Asymmetry maps are overplotted with Doppler velocity contour of –5 km s$^{-1}$ and magnetograms are overplotted with $\alpha_{\text{asym}}$ contour of 0.1. Asymmetry maps are saturated at 0.1. Notation in the first column corresponds to Figures 3 and 8. "Failed" refers to the failed eruption and "CME" to the CME event.

Nonthermal velocities of the deblended Fe XII 195.12 Å and unblended Fe XIII 202.04 Å lines were determined by

$$\delta\lambda = \frac{\lambda_0}{c}\sqrt{4\ln 2\left(\frac{2\,k_B T_i}{m}+\xi^2\right)+\sigma_I^2}, \quad (1)$$

where $\delta\lambda$ is the observed line width, $\lambda_0$ is the line centroid, $k_B$ is Boltzmann's constant, $T_i$ is the ion temperature, $m$ is the mass, $\xi$ is the nonthermal velocity, and $\sigma_I$ is the instrumental width (e.g. Brooks and Warren, 2016). The instrumental width of EIS varies for the 1″ and 2″ slits as well as the CCD $Y$-pixel number. The EIS SSW routine eis_slit_width.pro was used to account for these variations. Fe XII 195.12 Å intensity, Doppler and nonthermal velocity maps are shown in the first three columns of Figure 5. Each map is overplotted with MDI contours of $\pm 100$ G representing positive/negative (cyan/black) polarities. The EIS images have been cropped to 180″ in $Y$ to focus on the sigmoid. A similar set of Fe XIII 202.04 Å maps is displayed in Figure 6 without the MDI contours.

### 3.3. Blue-Wing Asymmetries - $\alpha_{asym}$

As mentioned in Section 1, earlier studies have shown that coronal upflows at the edges of active regions have intermittent and weak, yet noticeable enhancements in the blue wing of the line profiles. The presence of blue-wing asymmetries suggests that there are unresolved superposed high-speed plasma flows along the line of sight. Typically, the upflows are in the range of [50–100] km s$^{-1}$, however, they can be as high as 150 km s$^{-1}$. Identifying these subtle asymmetries can be problematic as they are only a few percent of line core intensities. One way is to fit double-Gaussian functions to the line profiles, however, this method assumes that the unresolved upflows are composed of a nearly stationary background component and a high-speed component. In this analysis, we do not assume a two-component fit to the profiles. Rather, we quantify the subtle asymmetry over the velocity range $[v_1 - v_2]$ km s$^{-1}$ by

$$\alpha_{asym} = \frac{A_{Obs} - A_{SG}}{(A_{Obs} + A_{SG})/2}, \quad (2)$$

where, in the velocity range $[v_1 - v_2]$ km s$^{-1}$, $A_{Obs}$ is the area under the observed intensity curve and $A_{SG}$ is the area under the fitted single Gaussian function. The value of $v_1$ is selected to remove most of the line core, while keeping the largest possible domain where the observed spectra is above the fitted Gaussian. The value of $v_2$ is selected to be as far as possible in the blue wing but set to avoid contamination of nearby emission lines. Initially, we experimented with fitting a single-Gaussian function to only the core intensity before calculating $\alpha_{asym}$ to assess the effect of the wings on the fit. In general, the fits were similar, and therefore we decided to use a single-Gaussian fit to the entire observed spectrum to calculate $\alpha_{asym}$ within the specified velocity range, which is simpler as the core range does not need to be defined. The method was applied to both the Fe XIII 202.04 Å and the Fe XII 195.12 Å emission lines.

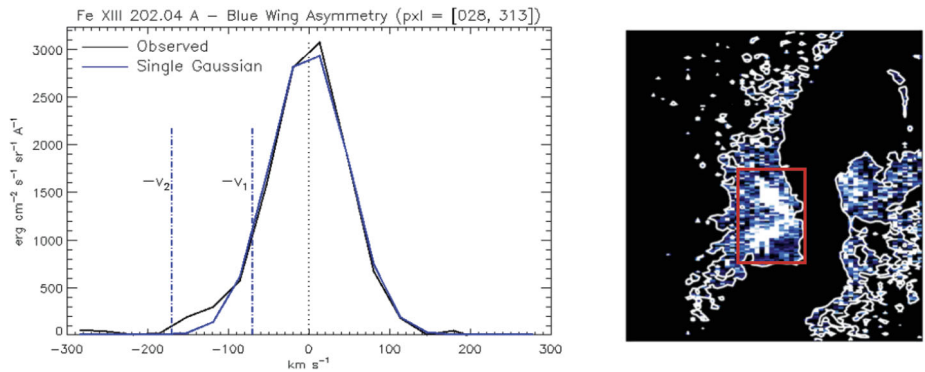

**Figure 7** Left: *Hinode*/EIS Fe XIII 202.04 Å observed emission line profile (black) overlaid with the single Gaussian fitted function (blue) for pixel = [28, 313] of the raster at 00:18 UT on 7 December 2007. The dotted line is the centroid of the fitted function and the dot-dash blue lines designate the velocity range $[v_1, v_2] = [-70, -170]$ km s$^{-1}$ over which the blue-wing asymmetry $\alpha_{\rm asym}$ is calculated. Right: Blue-wing asymmetry map at 00:18 UT. Spatial scale of the image is $X = 180''$, $Y = 180''$. An animation scanning the emission profiles for the pixels within the red box is included in the Supplementary Information as Fig6.mp4.

Figure 7 shows the Fe XIII 202.04 Å observed emission-line profile overlaid with the single-Gaussian fitted profile for a pixel in the eastern upflow region at 00:18 UT on 7 December 2007. The selected velocity range of [70 – 170] km s$^{-1}$ is delimited by the blue dot–dash lines. The velocity range $v_1$ to $v_2$ is of the order of the sound speed: 150 km s$^{-1}$ for a 1 MK plasma. Since the upflows are typically inclined away from the local vertical by an angle $\delta$, and this active region is near to the central meridian passage, the true plasma flows are larger by a factor of about $1/\cos\delta$. Baker et al. (2017) found $|\delta|$ to be in the range $[0° – 40°]$ for a set of ten active regions. Assuming AR 10977 is within this range, the plasma velocity can be [90 – 210] km s$^{-1}$, which is only a factor of 1.3 larger. We conclude that the observed flows have a Mach number around unity. A sample of blue-wing asymmetries can be found in the included animation of profiles for each pixel within the red box shown in the asymmetry map in the right panel of Figure 7.

Asymmetry maps are displayed in the fourth columns of Figures 5 and 6. Pixels outside of the upflow regions have been masked so that only asymmetries within the Doppler velocity contours of $-5$ km s$^{-1}$ are shown; outside of the contours, $\alpha_{\rm asym}$ is very noisy, without any coherent structure. The maps have been saturated at $\alpha_{\rm asym}$ of 0.05 for Fe XII and 0.10 for Fe XIII to emphasize the locations of the strongest asymmetries. It should be noted that small patches of blue-wing asymmetries were also observed in the upflow regions of AR 10977 before the raster at 12:03 UT on 6 December and after the one at 11:26 UT on 7 December (panels a and f in Figures 5 and 6). This raster was selected to represent typical cases outside of the eruptive period.

## 4. Upflow Evolution

### 4.1. Location and Evolution of Velocity Parameters

The *Hinode*/EIS Doppler velocity maps in Figures 5 and 6 show that the eastern/western upflow regions are located over the following/leading magnetic polarities on either side of the active region, i.e. in pairs (see the MDI contours overlaid on the Fe XII intensity, Doppler and nonthermal velocity maps in Figure 5). At 12:03 UT on 6 December, 13 hours prior to the failed eruption, upflows on the eastern side are concentrated in the northern end of the positive polarity near to the main PIL. However, on the western side, the spatial extent of the upflows is spread over patches of negative field farther from the PIL as the leading polarity is more fragmented compared to its positive counterpart (Figure 1). As the sigmoid/flux rope develops, the eastern upflow region expands but still appears to fan out from where it is rooted in the northern section of the positive field (00:18 UT on 7 December). The spatial extent of the blue-shifted plasma remains relatively large on both sides throughout the period of the failed eruption and CME (01:15, 03:27, and 06:37 UT) before significantly shrinking 6.5 hours after the CME (11:26 UT). Fe XII and Fe XIII upflows evolved in parallel throughout the observation period, which is not surprising given that the lines sample plasma at similar temperatures ($\approx 1.3 - 1.6$ MK; Young et al., 2007).

A parallel evolution of nonthermal velocities within the upflow regions is evident in the maps of Figures 5 and 6. At 12:03 UT on 6 December, before the period of eruptions, there are small patches of enhanced nonthermal velocities of $25-40$ km s$^{-1}$, mainly corresponding to the stronger magnetic-field regions on either side of the active region. The magnitude and locations of the nonthermal velocities are typical for quiescent active regions (e.g. Doschek et al., 2008). Nonthermal velocities become much stronger, exceeding 50 km s$^{-1}$, and more extended just after the peak times of each event before weakening and shrinking late in the decay phases. The evolution is similar for both emission lines.

Like the Doppler and nonthermal velocities, the evolution of the blue-wing asymmetries in the upflow regions appears to be related to the timeline of the failed eruption and CME especially on the eastern side. Long before the onset of the sigmoid/flux-rope activation, the asymmetry maps show small patches of enhanced $\alpha_{\rm asym}$ in the northern section of the positive polarity with more extended patches over the dispersed negative field (see right panels of Figures 5, 6). The area of strong blue-wing asymmetries is largest at the onset (00:18 UT) and immediately after the peak (01:15 UT) of the failed eruption. At 11:26 UT, 6.5 hours after the CME, the spatial extent shrinks to less than that of 13 hours before the eruptive period (top panel). Interestingly, the locations of the $\alpha_{\rm asym}$ contours shift from the inside to outside edge of the polarities away from the main PIL as the eruptions progress and the field surrounding the eruptive structure relaxes afterwards. See the red arrows in the last column of Figure 5 indicating the locations of the PIL and western edge of the $\alpha_{\rm asym}$ contours. Also, during the same time period an evolution from North to South is observed in the positive polarity, and in the reverse direction for the negative polarity. The evolution in the north–south direction is likely to be linked to the redistribution of the magnetic shear in the active region immediately following the eruption (Green, Kliem, and Wallace, 2011).

### 4.2. Global Evolution of Velocity Parameters

The global evolution of the three velocity parameters described above is quantified in Figure 8, which shows histograms of the Fe XIII Doppler and nonthermal velocities and $\alpha_{\rm asym}$ contained within the upflow region/contour on the eastern side of the active region for each raster in Figure 6 a–f. We have excluded the data for the western upflow region in view of the fact that it is cropped in the EIS maps, and importantly, cropping is time dependent. For the failed event (Figure 8 a–c, the frequency of strong Doppler velocities in the range of [–20, –10] km s$^{-1}$ increased prior to (00:18 UT) and again immediately after (01:15 UT) the peak of the failed eruption compared to the quiescent distribution 13 hours earlier. The nonthermal velocity distributions exhibited significant intensification for velocities greater than 25 km s$^{-1}$ over the same time period. Blue-wing asymmetries showed more modest increases up to an $\alpha_{\rm asym}$ of about 0.15 while remaining essentially unchanged for higher $\alpha_{\rm asym}$. Late in the decay phase at 03:27 UT, all three velocity parameters retreated to the levels/distributions similar to those observed at 00:18 UT at the onset of the failed event, but not fully back to quiescent levels from the previous day; the relaxation is evident in the upper tail of the distributions. Doppler velocities then exhibited a subtle enhancement in the tail below $-13$ km s$^{-1}$ 1.5 hours after the CME event before falling to below the pre-eruption quiescent levels by 11:26 UT (Figure 8 d). The entire nonthermal-velocity frequency distribution shifted toward substantially stronger values a few hours after the CME. Similarly, the intensification of $\alpha_{\rm asym}$ at 06:37 UT was markedly stronger: above 0.10 compared to just after the peak (01:15 UT) and during the decay phase (03:27 UT) of the failed event. All parameters returned to at/below quiescent/pre-eruption distributions by 11:26 UT.

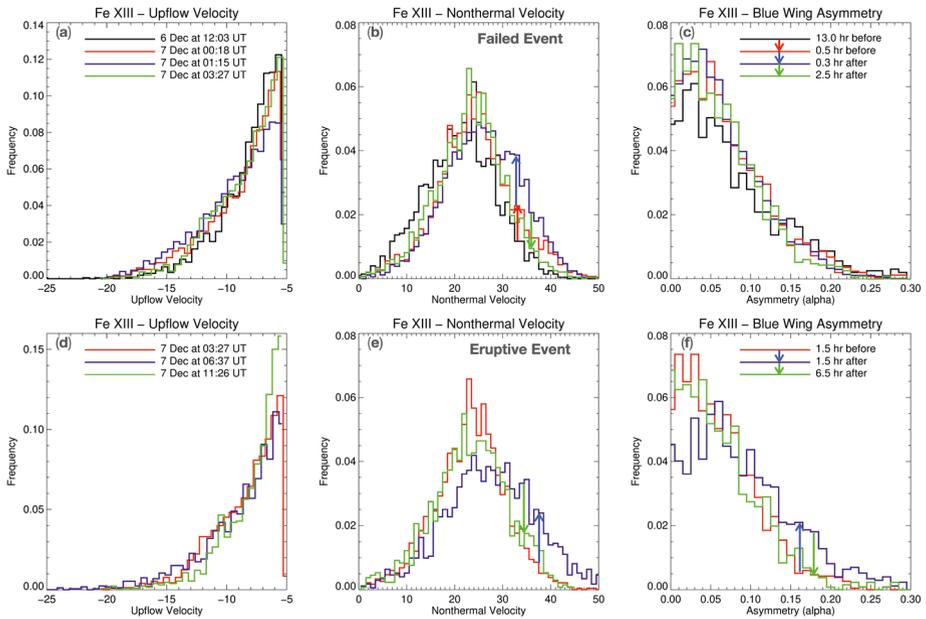

**Figure 8** Histograms of Fe XIII 202.04 Å Doppler upflow and nonthermal velocities, and blue-wing asymmetries, [$\alpha_{\rm asym}$], for the upflow region located on the eastern side of the active region for the failed (top panels) and eruptive events (bottom panels). The relative frequencies are computed within the upflow contours shown in Figure 6 (a–f). Arrows indicate the significant evolution of the nonthermal velocities and $\alpha_{\rm asym}$.

The similarity in the evolution of the velocity parameters suggests that they are correlated. In fact, we found low spatial correlation between $\alpha_{\rm asym}$ and the other parameters (the Pearson correlation coefficients for Doppler velocity are in the range = [0.02, 0.12] and for nonthermal velocity in the range = [0.03, 0.15]); however, there is a moderate to strong correlation between Doppler and nonthermal velocities. The Pearson correlation coefficients for Fe XIII Doppler and nonthermal velocities within the eastern upflow regions are given in Table 2. The correlation coefficient 13 hours before the eruptive period was $-0.40$ and increased to [$-0.54, -0.61$] during the failed eruption and CME. These results are in agreement with Doschek et al. (2008), where Doppler velocities in upflow regions of two active regions are confirmed to be correlated with nonthermal velocities.

## 5. Magnetic Reconnection Driven Upflows

The apparent link between the evolution of the upflows and the timeline of the eruptions suggests that the upflow driving mechanism, i.e. magnetic reconnection at specific topolog-ical locations, may be affected by or related to the formation and eruption of the flux rope. Although the observed enhancements in velocity parameters appear to be related in time to the flux-rope evolution, they are not co-spatial with the location of the flux rope as the upflow regions are at the sides of the observationally inferred flux-rope volume, i.e. on either side of the sigmoid. The key questions are: Where are the strongest upflows generated and what causes the intensification of upflows during the failed eruption and CME?

**Table 2** Pearson correlation coefficients for Doppler vs. nonthermal velocities in the eastern upflow region for *Hinode*/EIS rasters.

| Date | 6 Dec | 7 Dec | 7 Dec | 7 Dec | 7 Dec | 7 Dec |
|---|---|---|---|---|---|---|
| Time [UT] | 12:03 | 00:18 | 01:15 | 03:27 | 06:37 | 11:26 |
| Corr. coef. | −0.40 | −0.59 | −0.54 | −0.54 | −0.61 | −0.55 |

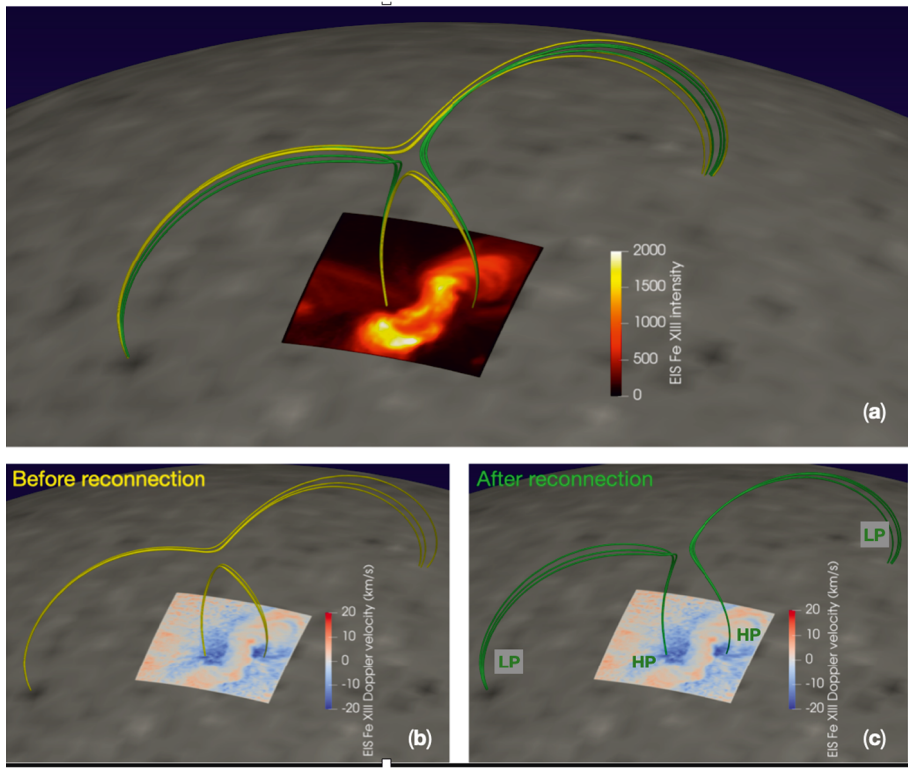

**Figure 9** PFSS extrapolation at 00:04 UT on 7 December 2007 showing a quadrupolar configuration with a magnetic null point located above the sigmoid (**a**). Field lines are plotted on the *Hinode*/EIS Fe XIII 202.04 Å intensity map at 00:18 UT and the background magnetogram (the intensity map is plotted as observed with no vertical extension). The selected field lines plotted on the corresponding Doppler velocity map are color-coded to indicate the magnetic configuration before (**b**, yellow) and after (**c**, green) magnetic reconnection. HP is high pressure and LP is low pressure as found in the loops just after reconnection. Note that in the EIS intensity map, the bright sigmoid dominates the much fainter arcade field. The extent of the central arcade field in panel b appears broad in this case because the twisted/sheared field of the sigmoid is not included in the potential field extrapolation (see related text).

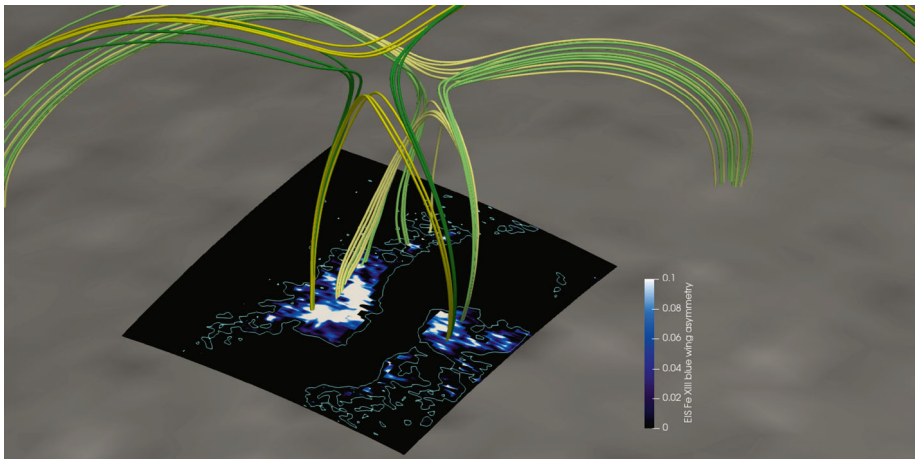

**Figure 10** PFSS extrapolation at 00:04 UT on 7 December 2007 plotted on the blue-wing asymmetry map at 00:18 UT (see Figure 6 b). $\alpha_{\text{asym}}$ is saturated at 0.1.

### 5.1. PFSS Extrapolation

In order to answer these questions, we make use of a potential field source surface (PFSS) extrapolation to investigate the large-scale coronal magnetic configuration in and around AR 10977 just before the failed eruption. The extrapolation of the photospheric magnetic field at 00:04 UT on 7 December 2007 was created using the IDL SolarSoft PFSS package (Schrijver and De Rosa, 2003) and is shown in Figure 9 a – c. Selected field lines are plotted on the EIS Fe XIII intensity (a) and Doppler velocity maps (b, c) at 00:18 UT and the surrounding photospheric magnetogram. The sigmoid is not included in the extrapolation as its field is highly sheared/twisted and therefore non-potential (Aulanier et al., 2010; Green, Kliem, and Wallace, 2011). The aim of this analysis is to examine the large-scale magnetic field computed with observed boundary conditions. Such a large scale field is expected to be much closer to a potential field than the sigmoid core region.

### 5.2. Magnetic Topology

The global magnetic topology is a quadrupolar configuration with a magnetic null point above the sigmoid/flux rope. Four distinct connectivities of field lines are present in the extrapolation (Figure 9 a): i) arcade field lines (yellow) located above the sigmoid and rooted within the active region; ii) long field lines (yellow) lying above the arcade field connecting weak magnetic fields away from the active region; iii) field lines (green) rooted on one end in the positive polarity of the upflow regions and on the other end in the weaker negative field to the Southeast of the active region; iv) field lines (green) rooted on one end in the negative polarity of the upflow regions and on the other end in the weaker positive field to the Northwest of the active region. The null point is at the central position where the four types of connectivities are the closest in Figure 9 a. It should be noted that in the EIS intensity map of Figure 9 and in the EUV images of STEREO-A in Figure 2, we mainly see the much brighter sigmoid, not the overlying arcade field. The sigmoid field contains large currents and undergoes internal reconnection, generating more heating compared to the arcade field.

In the null-point configuration, as the sigmoid/flux rope expands and rises during the building phase (photospheric cancelation), then in the failed eruption and CME, the arcade above it also is forced to rise up, pushing against the overlying long magnetic-field lines (yellow field lines in Figure 9 b). This drives reconnection at the null point, creating the new magnetic-field lines in green in Figure 9 c. The above description is mostly a 2D description, as the field lines passing near the null point are almost contained in a plane. A sense of the 3D nature of the configuration is shown in Figure 10 with four other sets of similar field lines drawn away from the null point (pale yellow and green colors). Their closest approach outlines a so-called separator, which is a curve that separates the four types of magnetic connectivities. The separator could also be viewed as the intersection of two dome-shaped separatrix surfaces. Each separatrix delimits the magnetic connectivities linked to the leading/following magnetic polarity of the active region. The intersecting separatrices define four spatial regions, or domains, where the magnetic connectivities are similar to the example shown in Figure 10.

### 5.3. Relationship with Strong $\alpha_{\text{asym}}$

In Figure 10, selected field lines are plotted on an Fe XIII blue-wing asymmetry [$\alpha_{\text{asym}}$] map at 00:18 UT. The PFSS extrapolation shows that the green reconnected field lines are located in close proximity to the strongest $\alpha_{\text{asym}}$ values on both sides of the active region. An exact match is not to be expected due to the limitations of the PFSS extrapolation method (e.g. Wiegelmann and Sakurai, 2012). Moreover, a synoptic magnetogram is used in the extrapolation so that even if it is updated, there is still a step evolution of six hours. On the other hand, the large-scale photospheric configuration that determines the overall null-point structure is evolving on an active-region time scale that is much longer than the timeline of events considered here. From a structural point of view, the quadrupolar configuration in Figure 9 has a very stable magnetic topology where the separatrices are only shifted in position by the inclusion of magnetic shear (e.g. Démoulin, 2006). On the eastern side, the elongated shape of strong $\alpha_{\text{asym}}$ traces the footprint of the separatrix. On the western side, the area of strong $\alpha_{\text{asym}}$ is more compact because the negative polarity is more concentrated. Analogous to H$\alpha$ flare ribbons tracing the intersection of separatrices with the photosphere/chromosphere (e.g. Démoulin, 2006), the pattern of strong $\alpha_{\text{asym}}$ follows the extent of separatrices in the low corona.

Since the PFSS extrapolation is a static computation of the coronal field, it cannot show evolution due to reconnection. However, the extrapolation can show the approximate locations of the stable separatrices, thereby identifying where reconnection is expected to take place. In the case of AR 10977, the evolution is mainly driven by the active-region core. If a sheared/twisted core were included in the extrapolation, the additional induced magnetic pressure would push the yellow arcade further upward to reconnect in resistive MHD. In a fully nonlinear model, we would expect the separatrices to be closer to the main PIL as the shear/twist is stronger in the active-region core. We conclude that the placement of the separatrix footprint to the outside of the traces of strong $\alpha$, rather than at the periphery of the AR core loops, can be fully ascribed to the limitation of the coronal model.

We looked for other indications of reconnection in this quadrupolar configuration. The distant footpoints of field lines passing close to the separator (i.e. the far footpoints of the green field lines away from the active region in Figure 9 a) define the regions where brightenings are expected as the released energy is mostly transported by thermal conduction and accelerated particles along field lines. This method has provided the means to find faint kernels in flares (e.g. Bagalá et al., 1995). Here we did not find any brightenings related to the quadrupolar reconnection, which is not surprising since the energy released is much weaker than in a flare.

Given that the location and evolution of blue-wing asymmetries appear to be directly linked to the magnetic topology, they provide additional information about the underlying physics of upflows over the other velocity parameters. In effect, the blue-wing asymmetries are a tracer of the *most recent reconnection* in the loops rooted in the upflow regions. In previous studies, this signal was washed out in the Doppler and nonthermal velocities since only a tiny fraction of the plasma in the upflows is observed just after reconnection, before it expands significantly and the speed decreases. Even in the pixels where $\alpha_{\text{asym}}$ is large, there is only a small fraction of plasma emitting in the far blue wing compared to the rest of the observed profile (Figure 7). The detection of plasma upflowing just after reconnection requires defining a dedicated parameter such as $\alpha_{\text{asym}}$.

### 5.4. Evolution of Failed Eruption and CME

Next, we analyze the temporal evolution of upflows in the context of the magnetic topology. At 00:18 UT, close to the time of the PFSS extrapolation, an increase in the three velocity parameters is already evident (Figure 8), suggesting that quadrupolar reconnection is at work at least 30 minutes before the failed eruption. The velocity parameters continue to strengthen at 01:15 UT, 20 minutes after the peak of the failed eruption. During this time period, the sigmoid continues to expand and rise (see included animation of Figure 2), inducing more reconnection along the separator lying above it. Reconnection is likely to continue as the sigmoid rises until it reaches a new equilibrium height. The animation of Figure 2 shows that the sigmoid appears to pause in its rise phase between 03:06 – 04:06 UT, however, the low temporal cadence prevents more definitive confirmation. Relaxation is present at 03:27 UT during the decay phase of the failed event when the velocity parameters have returned close to pre-eruption quiescent distributions.

The cycle repeats when the system erupts in a catastrophic way, leading to a CME. There is a significant increase in the velocity parameters once again for the histogram distributions at 06:37 UT, 90 minutes after the peak of the CME. The system has relaxed by the time of the next EIS observation at 11:26 UT. A notable difference between the two eruption cycles is that the increase in blue-wing asymmetries during the CME event is significantly more enhanced compared to that leading up to the failed eruption.

### 5.5. Summary of Evolution

In summary, the location and evolution of the velocity parameters fit very well the scenario of separator reconnection that is amplified during the eruptions. The amplification of reconnection is directly related to how much the magnetic configuration is transformed as the flux rope moves through successive stages – from the relatively quiescent early stages of flux-rope formation to the failed eruption and finally to the CME. Evolution of the velocity parameters, especially nonthermal velocities and blue-wing asymmetries, reflects the cycle of slow and fast expansion of the system throughout the evolution of the flux rope. However, notably the upflows and in particular the blue-wing asymmetries do not stop during the quiet phases, demonstrating that quadrupolar reconnection is still ongoing. This is an indication that the observed photospheric evolution during these phases does not build a dissipationless current sheet along the separator, as in ideal MHD, but rather reconnection is continuously at work. This result is in contrast to the breakout model, which postulates no reconnection so that a long current sheet forms around the null point while continuous shearing at the base of the underlying arcade increases the free energy of the system. Once the current sheet becomes unstable, then the quadrupolar reconnection takes place and the core field is released from below (Antiochos, DeVore, and Klimchuk, 1999; Lynch et al., 2008).

Ideally, a much higher temporal cadence is needed to analyze the evolution of $\alpha_{asym}$ before and during an eruption. Since it measures the emission of plasma accelerated in the blue wing, normalized by the emission of the line core, $\alpha_{asym}$ is related to the amount of magnetic flux reconnected per unit time. This is modulated by other quantities such as the pressure difference between the reconnected loops (see Section 5.6) and the amount of heating released, so the induced ionization level. If the evolution of these parameters can be derived from observations, monitoring $\alpha_{asym}$ provides information about the reconnection-rate evolution.

### 5.6. Driving Mechanism of Upflows

The yellow arcade loops in the active region are dense and hot so that they are clearly seen in the intensity maps of Figures 5 and 6. In contrast, we could not find any trace of the long yellow field lines in EUV images. This is not surprising since they are located in much weaker magnetic field; the loops are cooler and more tenuous than the arcade loops. Consequently, the two sets of connectivities contain plasma with drastically different thermal pressures. The pressure imbalance is maintained because of the low plasma-$\beta$ of the corona, however, once reconnection takes place, this is no longer the case.

The new flux tubes formed by reconnection are in a state of strong over-pressure for the section closest to the active region compared to the other section rooted in the weak field (see green field lines and notations in Figure 9 c). In this framework, reconnection is like opening a floodgate to release the plasma contained in the arcade flux tube into an almost empty large volume. The post-reconnection pressure gradient drives strong upflows along the reconnected flux tubes at a speed of the order of the sound speed: $\approx 150$ km s$^{-1}$ for a plasma at 1.0 MK. This is consistent with the blue-wing velocities observed in AR 10977 and those cited in Section 1. It is also supported by the numerical simulations of Bradshaw, Aulanier, and Del Zanna (2011).

### 6. Conclusion

In this study, we analyze the evolution of active-region upflows in AR 10977 before, during, and after a failed eruption of a flux rope followed by a full eruption of the flux rope as a CME. Three velocity parameters – Doppler and nonthermal velocities and blue-wing asymmetries – increased during the onset and then relaxed to quiescent finite levels during the decay phases of both eruptions. Significant enhancements in nonthermal velocities and blue-wing asymmetries were observed around the time of the CME. A PFSS extrapolation of the active region contains a quadrupolar configuration with a separator lying above the flux rope. Such a configuration is consistent with a scenario where, as the flux rope expands and rises, the dense, hot arcade loops, located above the flux rope, are pushed into the cooler extended overlying loops, inducing reconnection along the separator and thereby creating a significant pressure gradient in the newly formed loops. The post-reconnection pressure gradient drives strong upflows along the field lines that are spatially linked to the blue-wing asymmetries within the upflow regions; the more recent the reconnection, the higher the asymmetry. Furthermore, the evolution of the asymmetries closely follows the cycle of expansion and relaxation of the active-region core.

In the low plasma-$\beta$ corona, magnetic pressure is significantly higher than plasma pressure. Although the corona is dominated by magnetic field, the gradient of plasma pressure remains a powerful driver of plasma flows along the magnetic field on all scales. Here we have demonstrated the role of the pressure gradient in driving the strongest active-region upflows in quiet and eruptive periods in AR 10977.

Magnetic configurations with complex topologies that include separatrices, e.g. null points (e.g. Del Zanna et al., 2011) and separators, have been generalized in 3D to quasi-separatrix layers or QSLs (Démoulin et al., 1996). This means that AR 10977 provides a clear case of QSL reconnection creating and driving active-region upflows as first proposed by Baker et al. (2009) and followed by van Driel-Gesztelyi et al. (2012), Démoulin et al. (2013), Mandrini et al. (2015), and Baker et al. (2017). The global evolution of the velocity parameters is more extreme in AR 10977 than what is typically found in quiescent active regions; it is a matter of scale with the failed eruption and CME forcing more reconnection, which in turn causes stronger flows over a more compressed time period of hours rather than days/weeks. The fact that the blue-wing asymmetries are present days before and hours after the eruptions suggests that its driving mechanism, i.e. QSL reconnection, is always at work in active regions. This is supported by the results of Démoulin et al. (2013) and Baker et al. (2017) who showed that the upflow velocities are persistent on both the following and leading AR polarities, with a similar inclination to the local vertical seen hardly to change as active regions cross the solar disk.

The augmented observational capabilities available with spacecraft positioned at different positions will further enhance our understanding of the nature of active-region upflows and the processes that create and drive them. Spectroscopically, the coordination between *Hinode*/EIS, IRIS, and *Solar Orbiter*/*Spectral Imaging of the Coronal Environment* (SPICE) in interesting configurations such as the many quadratures offered during the lifetime of the *Solar Orbiter* mission, will provide new insight into the evolution and characteristics of these upflows. Furthermore, magnetic-field measurements in similar configurations from different viewing angles such as provided by, e.g., SDO/*Helioseismic and Magnetic Imager* (HMI) and *Solar Orbiter*/*Polarimetric and Helioseismic Imager* (PHI), will advance our understanding of the magnetic-field geometry of the active regions by better constraining extrapolations. This will help us to further advance our understanding of the formation and drivers of active-region upflows.

**Supplementary Information** The online version contains supplementary material available at https://doi.org/10.1007/s11207-021-01849-7.

**Acknowledgements** *Hinode* is a Japanese mission developed and launched by ISAS/JAXA, collaborating with NAOJ as a domestic partner, and NASA and STFC (UK) as international partners. Scientific operation of *Hinode* is performed by the *Hinode* science team organized at ISAS/JAXA. This team mainly consists of scientists from institutes in the partner countries. Support for the post-launch operation is provided by JAXA and NAOJ (Japan), STFC (UK), NASA, ESA, and NSC (Norway). D. Baker is funded under STFC consolidated grant number ST/S000240/1 and L. van Driel-Gesztelyi is partially funded under the same grant and acknowledges the Hungarian National Research, Development and Innovation Office grant OTKA K-131508. G. Valori acknowledges the support from the European Union's Horizon 2020 research and innovation programme under grant agreement No 824135 and of the STFC grant number ST/T000317/1. The work of D.H. Brooks was performed under contract to the Naval Research Laboratory and was funded by the NASA *Hinode* program. D.M. Long is grateful to the Science Technology and Facilities Council for the award of an Ernest Rutherford Fellowship (ST/R003246/1). We recognise the collaborative and open nature of knowledge creation and dissemination, under the control of the academic community.

## Declarations

**Disclosure of Potential Conflicts of Interest** The authors declare that they have no conflicts of interest.